\documentclass[aps,prb,twocolumn,groupedaddress,showpacs,amsmath,amssymb]{revtex4}

\usepackage{graphicx}
\usepackage{dcolumn}
\usepackage{bm}

\begin{document}

\title{Fermi level tuning and a large activation gap achieved in the
topological insulator Bi$_{2}$Te$_{2}$Se by Sn doping}

\author{Zhi Ren}
\author{A. A. Taskin}
\author{Satoshi Sasaki}
\author{Kouji Segawa}
\author{Yoichi Ando}
\email{y_ando@sanken.osaka-u.ac.jp}

\affiliation{Institute of Scientific and Industrial Research,
Osaka University, Ibaraki, Osaka 567-0047, Japan}

\date{\today}

\begin{abstract}
We report the effect of Sn doping on the transport properties of the
topological insulator Bi$_{2}$Te$_{2}$Se studied in a series of
Bi$_{2-x}$Sn$_{x}$Te$_{2}$Se crystals with 0 $\leq$ $x$ $\leq$ 0.02. The
undoped stoichiometric compound ($x$ = 0) shows an $n$-type metallic
behavior with its Fermi level pinned to the conduction band. In the
doped compound, it is found that Sn acts as an acceptor and leads to a
downshift of the Fermi level. For $x$ $\geq$ 0.004, the Fermi level is
lowered into the bulk forbidden gap and the crystals present a
resistivity considerably larger than 1 $\Omega$cm at low temperatures.
In those crystals, the high-temperature transport properties are
essentially governed by thermally-activated carriers whose activation
energy is 95--125 meV, which probably signifies the formation of a
Sn-related impurity band. In addition, the surface conductance directly
obtained from the Shubnikov-de Haas oscillations indicates that a
surface-dominated transport can be achieved in samples with several $\mu$m
thickness.
\end{abstract}


%

\maketitle

\section{Introduction}

In a three-dimensional (3D) topological insulator (TI), a band inversion
due to strong spin-orbit coupling induces gapless surface states (SS)
consisting of spin helical Dirac Fermions,\cite{K1,MB,Roy,K2,Qi} which
are expected to give rise to a number of topological quantum phenomena.
\cite{Kane,ZhangSC} Whereas such surface sensitive probes as
angle-resolved photoemission spectroscopy (ARPES) and scanning tunneling
microscope have been successfully applied to investigate the topological
SS,\cite{ARPESHsieh,ARPESmatsuda,ARPESshen,ARPESSato,ARPESKuroda,
ARPESShen2,ARPESBTS,ARPESBSTS,STM1,STM2,STM3,STM4} the transport study
of the SS remains a challenge due to the presence of parallel bulk
conducting channel that usually dominates the transport
properties.\cite{Taskin,Ong2010,Fisher_np,HorBi2Se3,checkelsky,Butch,
Analytis,Eto,RenBi2Se3} Recently, the ternary tetradymite
Bi$_{2}$Te$_{2}$Se (BTS) has become a prototype TI material for studying
the peculiar spin and charge transport of the SS,\cite{RenBTS,OngBTS}
because carefully prepared BTS crystals not only present a very low bulk
conduction, but also show clear Shubnikov-de Haas (SdH) oscillations
that reflect a high mobility of surface Dirac electrons. Given the
lack of a truly bulk-insulating state in any of the known TI materials
so far,\cite{RenBSTS,BSTStaskin,HgTe} further improvement of BTS is worth
pursuing.

The first synthesis of BTS was made while optimizing the continuous
solid solutions between Bi$_{2}$Te$_{3}$ and Bi$_{2}$Se$_{3}$ for
thermoelectric applications.\cite{Bi2Te3-xSex} Nevertheless, unlike other
members in the Bi$_{2}$Te$_{3-x}$Se$_{x}$ (0$<$$x$$<$3) family, the Te
and Se atoms in BTS occupy distinct crystallographic sites, forming
quintuple layers arranged in the sequential order Te-Bi-Se-Bi-Te along
the $c$ axis.\cite{BSTSordering} Such chalcogen ordering is believed to
provide structural basis for reducing bulk carriers and achieving high
surface mobility.\cite{RenBTS} Unfortunately, similar to
Bi$_{2}$Se$_{3}$, the BTS crystals grown from the stoichiometric melts
show $n$-type metallic conduction with the electron density $n_{e}$ of
$\sim$10$^{19}$ cm$^{-3}$.\cite{JiaBTS} When the
chalcogen stoichiometry is altered in the starting composition, one
obtains\cite{RenBTS,JiaBTS}
metallic or insulating crystals at different positions along the
boule, which is due to the inevitable phase separation during the
solidification process, according to a newly established phase diagram
of the Bi$_{2}$Te$_{3}$--Bi$_{2}$Se$_{3}$ system.\cite{BTScrystal}
Therefore, it is desirable to explore an alternative route to obtain BTS
crystals with a large bulk resistivity while keeping the chalcogen
stoichiometry, which would be useful for achieving a high surface-carrier
mobility.

In this paper, we show that the above objective can be accomplished by
hole doping through a Bi-site substitution, which has already been
proven to be effective in tuning the carrier type and density for both
of the binary end members, Bi$_{2}$Te$_{3}$ and
Bi$_{2}$Se$_{3}$.\cite{ARPESshen,Fisher_np,HorBi2Se3,RenBi2Se3} It is
found that, in analogy to its role in
Bi$_{2}$Te$_{3}$,\cite{ARPESshen,HeavilySnBTS,SnBTS-Kulbachiskii,
SnBTS-resonant} the group IV element Sn acts as an acceptor in BTS,
which allows us to tune the Fermi level ($E_{\rm F}$) of this material.
In particular, the Bi$_{2-x}$Sn$_{x}$Te$_{2}$Se crystals with $x$ $\geq$
0.004 show low-temperature resistivity reaching several $\Omega$cm,
indicating that $E_{\rm F}$ is tuned into the bulk band gap; their
transport properties at high temperatures signify a large activation
gap, which probably comes from a Sn-related impurity band. In about a
half of those samples, we observed SdH oscillations originating from the
topological surface state below 14 T. Our analysis of the SdH
oscillations gives direct evidence that one can achieve a
surface-dominated transport in a bulk Sn-doped BTS crystal with a
thickness of several $\mu$m.

\section{Experimental Details}

The single crystals of Bi$_{2-x}$Sn$_{x}$Te$_{2}$Se were grown by
melting high purity elemental shots of Bi (99.9999\%), Sn (99.99\%), Te
(99.9999\%), and Se (99.999\%) with a nominal ratio of Bi:Sn:Te:Se =
(2-$x$):$x$:2:1 ($x$ = 0, 0.002, 0.004, 0.006, 0.01, 0.02) in sealed
evacuated quartz tubes at 850 $^{\circ}$C for 48 h with periodically
shaking to ensure homogeneity, followed by cooling slowly to 500
$^{\circ}$C and then annealing at that temperature for 4
days. The resulting crystals are easily cleaved along the (111) plane,
revealing a shiny mirrorlike surface. The x-ray diffraction (XRD)
analysis, which was performed on powders obtained by crushing the
crystals, confirmed the samples to be single phase with
chalcogen-ordered tetradymite structure.

For transport characterizations, the crystals were cut into bar-shaped
samples with the typical thickness of 100 $\mu$m after they were checked
to be single domain by x-ray Laue analysis. The electrical leads were
attached to the samples using room-temperature-cured silver paste in a
six-probe configuration. The in-plane resistivity $\rho_{xx}$ and the
Hall coefficient $R_{\rm H}$ were measured in a Quantum Design Physical
Properties Measurement System (PPMS-9) down to 1.8 K. The magnetic field
was applied along the $C_{3}$ axis which is perpendicular to the cleaved
surface. For each $x$ value, the data were taken on several crystals
obtained from different parts of the boule in order to check for
sample-to-sample variation. In addition, selected samples were cleaved
down to a few micrometers by using Kapton tapes and then brought to a
14-T magnet for the detection of SdH oscillations using an ac
measurement technique, in which two lock-in amplifiers were employed to
collect the signals in both the longitudinal ($\rho_{xx}$) and
transverse ($\rho_{yx}$) channels simultaneously. The measurements were
carried out by sweeping the magnetic field between $\pm$14 T at the
rate of 0.3 T/min, during which the temperature was stabilized to within
$\pm$5 mK.

\section{Results}

\subsection{Resistivity}

Figure 1(a) shows typical data for the temperature dependences of
$\rho_{xx}$ in Bi$_{2-x}$Sn$_{x}$Te$_{2}$Se crystals with different $x$
values, together with the data for an insulating sample grown from the
starting composition Bi$_{2}$Te$_{1.95}$Se$_{1.05}$ which we abbreviate
as BTS$_{1.05}$. The undoped stoichiometric BTS crystal ($x$ = 0) shows
a metallic behavior in the whole temperature range; correspondingly, the
Hall coefficient $R_{\rm H}$ [inset of Fig. 2(a)] is negative and nearly
temperature independent, giving an electron carrier density of
$\sim$1.5$\times$10$^{19}$ cm$^{-3}$ at 1.8 K. This result is in
agreement with the previous study,\cite{JiaBTS} which indicated that
$E_{\rm F}$ is pinned to the conduction band for this composition. Upon
doping with Sn, a drastic change in both the resistivity value and its
temperature dependence was observed. It is worth noting that an
insulating behavior of $\rho_{xx}$ is already established even for 0.1\%
of Sn doping to the Bi site ($x$ = 0.002). For $x$ $\geq$ 0.004, the
$\rho_{xx}$ values reach a few $\Omega$cm, which are usually larger than
that achieved in BTS$_{1.05}$.\cite{RenBTS}

\begin{figure}
\includegraphics*[width=7.5cm]{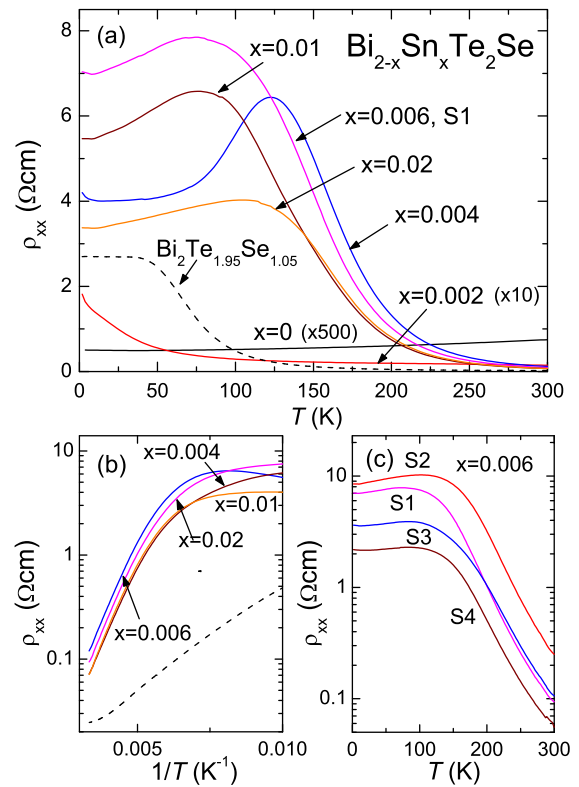}
\caption{(Color online)
(a) Temperature dependences of $\rho_{xx}$ for
Bi$_{2-x}$Sn$_{x}$Te$_{2}$Se single-crystal samples with various $x$.
Note that the data with $x$ = 0 and $x$ = 0.002 have been magnified by a
factor of 500 and 10, respectively. For comparison, the data for an
insulating sample grown from the starting composition
Bi$_{2}$Te$_{1.95}$Se$_{1.05}$ is also included (dashed line). (b)
Arrhenius plot of the $\rho_{xx}(T)$ data for the temperature range
between 100 and 300 K. (c) $\rho_{xx}(T)$ data for four representative
samples with $x$ = 0.006 obtained from different parts of a boule. The
data are labeled in numerical order, where S1 and S4 represent
first-to-freeze and the end parts, respectively. Note that the vertical
axis is in logarithmic scale. }
\label{fig1}
\end{figure}

It should be noted that there is a marked difference in the
$\rho_{xx}$($T$) behavior between the Sn-doped BTS ($x$ $\geq$ 0.004)
and the BTS$_{1.05}$ samples. Firstly, in the high-temperature region,
$\rho_{xx}$ in Sn-doped BTS samples increases more steeply with
decreasing temperature, as can be seen more clearly in the Arrhenius
plot shown in Fig. 1(b). The magnitude of $\rho_{xx}$ at room
temperature in Sn-doped BTS samples varies from 80 to 110 m$\Omega$cm,
which are considerably larger than that in BTS$_{1.05}$.\cite{RenBTS}
Also, the high-temperature slope in the Arrhenius plot, which is almost
doping independent, corresponds to an activation gap $\Delta$ of
approximately 120 meV; this is nearly three times larger than that found
for the BTS$_{1.05}$ sample ($\Delta$ $\simeq$ 45 meV). Secondly, in the
low temperature region, while $\rho_{xx}(T)$ of the BTS$_{1.05}$ sample
becomes nearly flat, the Sn-doped BTS samples show a more strongly
temperature-dependent $\rho_{xx}(T)$ behavior with a weak upturn below
$\sim$10 K, whose origin is not clear at present.

In addition to the above observations, the Sn-doped BTS samples were
found to show a higher degree of homogeneity. As an example, Fig. 1(c)
shows the $\rho_{xx}(T)$ data for four samples obtained from different
parts of a boule for $x$ = 0.006. One can see that all the samples show
qualitatively similar temperature dependence of $\rho_{xx}$. Although
the $\rho_{xx}$ values still vary between samples by a factor of 4, it
is notable that all of them are larger than 1 $\Omega$cm at low
temperature.

\subsection{Hall coefficient}

\begin{figure}
\includegraphics*[width=8cm]{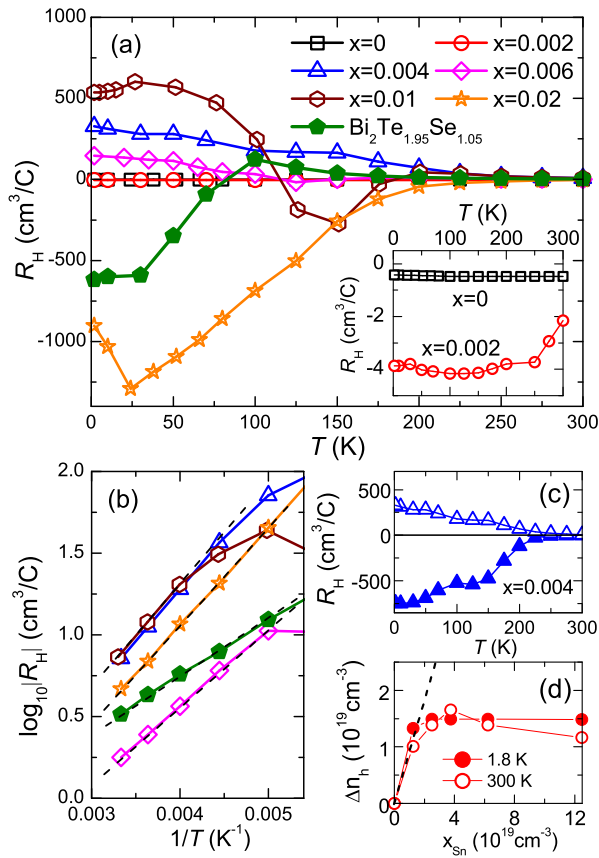}
\caption{(Color online)
(a) Temperature dependences of the low-field Hall coefficient $R_{\rm
H}$ for a series of Bi$_{2-x}$Sn$_{x}$Te$_{2}$Se samples, together with
the data for the BTS$_{1.05}$ sample. The inset shows a magnified view
of the data with $x$ = 0 and $x$ = 0.002. (b) Arrhenius plot of $R_{\rm
H}(T)$ at high temperature for a series of samples except for $x$
= 0 and 0.002 where no activation behavior was observed. Dashed
lines are the Arrhenius-law fittings to extract the activation energy
$\Delta^{\ast}$. (c) $R_{\rm H}(T)$ data for two different samples with
$x$ = 0.004, showing both $n$-type and $p$-type behavior. (d)
The density of introduced holes ($\Delta$$n_{h}$ = $n$($x$) - $n$(0)) by
Sn doping plotted as a function of the density of Sn dopant atoms. The
estimation of the carrier density is based on a one-band model for
simplicity. The dashed line corresponds to the ideal situation where
each Sn atom donates one hole. }
\label{fig2}
\end{figure}

To further investigate the doping effect of Sn in BTS, we measured the
Hall resistivity $\rho_{yx}$. Figure 2(a) shows the temperature
dependences of the Hall coefficient $R_{\rm H}$ for the same series of
BTS$_{1.05}$ and Bi$_{2-x}$Sn$_{x}$Te$_{2}$Se samples; here we follow
our previous definition of $R_{\rm H}$ as $R_{\rm H}$ = $\rho_{yx}/B$
near $B$ = 0.\cite{RenBSTS} It turns out that 0.1\% of Sn doping results
in an increase in $|R_{\rm H}|$ by nearly one order of magnitude, to
$\sim -4$ cm$^{3}$/C at 1.8 K [Fig. 2(a) inset]. Assuming a one-band
model, this gives $n_{\rm e} \simeq$ 1.6$\times$10$^{18}$ cm$^{-3}$,
implying that one Sn atom substituted for a Bi atom in BTS introduces
approximately one hole, which is much the same as that in the case of
$n$-type Bi$_{2}$Te$_{3}$.\cite{ARPESshen} While $R_{\rm H}$ is negative
for $x$ = 0 and 0.002 [Fig. 2(a) inset], it becomes positive at $x$ =
0.004 [Fig. 2(a) main panel], suggesting that an n-to-p type transition
occurs as a result of Sn doping between $x$ = 0.002 and 0.004. In fact,
in the case of $x$ = 0.004, both $n$- and $p$-type samples with a large
$|R_{\rm H}|$ were found in the same batch [Fig. 2(c)], placing this
composition at the verge of such a transition. Somewhat unexpectedly,
$R_{\rm H}$ becomes negative again at a higher doping of $x$ = 0.02
[Fig. 2(a)], whose origin is not clear. In the following, we focus on
the temperature dependence of $R_{\rm H}$ for the samples with $x$
$\geq$ 0.004.

Above $\sim$200 K, $R_{\rm H}$ of those samples shows a thermally
activated behavior, indicating that $E_{\rm F}$ is lowered into the bulk
band gap. The effective activation gap $\Delta^{\ast}$, which is
obtained from the Arrhenius plot of $R_{\rm H}(T)$ shown in Fig. 2(b),
is 115, 95, 125, and 110 meV for the samples with $x$ = 0.004, 0.006,
0.01, 0.02, respectively. Actually, the $\Delta^{\ast}$ values for all
the measured crystals fall within 95--125 meV for those compositions.
These $\Delta^{\ast}$ values are not far from the $\Delta$ values
derived from the $\rho_{xx}(T)$ data, and their difference is likely due
to the temperature dependence of the carrier mobility. More importantly,
the $\Delta^{\ast}$ values are much larger than that found in the
BTS$_{1.05}$ sample ($\sim$65 meV), implying that the Sn doping brings
$E_{F}$ closer to the middle of the band gap. At low temperatures, the
$|R_{\rm H}|$ values of those Sn-doped samples become large, sometimes
exceeding 1000 cm$^{3}$/C which would correspond to the carrier density
of only 6 $\times$ 10$^{15}$ cm$^{-3}$ in a one-band model.

\begin{figure}
\includegraphics*[width=8cm]{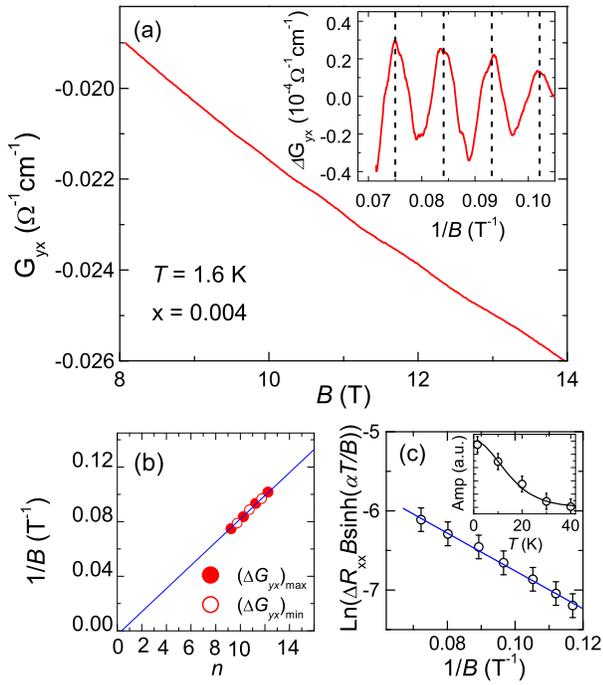}
\caption{(Color online)
(a) Transverse conductance $G_{yx}$
for an $n$-type sample with $x$ = 0.004 at 1.6 K plotted as a function of
magnetic field $B$ applied along the $C_{\rm 3}$ axis.
The inset shows the oscillatory component of $G_{yx}$,
$\Delta$$G_{yx}$, plotted as a function of 1/$B$. The
dashed lines are the guide to the eyes.
(b) Landau-level fan diagram for
oscillations in $\Delta$$G_{yx}$. Maxima and minima correspond to $n +
1/4$ and $n + 3/4$, respectively. The linear fitting with the slope of
$F$ = 116 T fixed by the oscillation frequency intersects the axis at
$\beta$ = 0.4 $\pm$ 0.1.
(c) Dingle plot for the oscillations in
$\Delta$$R_{xx}$, yielding $T_{D}$ = 12.5 K; inset shows the temperature
dependence of the SdH amplitudes. The solid line represents the fitting
by the LK theory, giving $m_{c}$ = 0.13$m_{e}$. }
\label{fig3}
\end{figure}

\subsection{Surface quantum oscillations}

Although Sn doping is expected to introduce impurity scattering, it
turned out that the surface mobility remains reasonably high so that SdH
oscillations can still be observed. As a matter of fact, traces of SdH
oscillations were detected in nearly 50\% of the Sn-doped BTS samples
measured in the 14-T magnet. Among those successful cases, the data
taken on an $n$-type sample with $x$ = 0.004 (thickness $t$ = 6 $\mu$m)
\cite{footnote} showed the simplest pattern of the oscillations. In the
following, we present the analysis of this simplest case.

Figure 3(a) shows the magnetic-field dependence of the transverse
conductivity $G_{yx}$ of this sample at 1.6 K, which was calculated from
$\rho_{xx}$ and $\rho_{yx}$. SdH oscillations are already visible in the
raw data for magnetic field above 10 T. After removing a smooth
background, one can clearly see that the oscillatory part of $
G_{yx}$, $\Delta$$G_{yx}$, exhibits periodic maxima and minima as a function
of $1/B$ [Fig.
3(a) inset], establishing the existence of a well defined Fermi surface
(FS). In Fig. 3(b), we plot the $1/B$ values corresponding to the maxima
(closed circles) and the minima (open circles) of $\Delta$$G_{yx}$ as a
function of the Landau level index $n$, following the index assignment
scheme in Ref. \onlinecite{BTSberryphase}. From the linear fitting of
the data with the slope fixed at the oscillation frequency obtained from
the Fourier transform ($F$ = 116 T), we obtain a finite intercept
$\beta$ = 0.4 $\pm$ 0.1. Since the slope is fixed in this analysis, the
error in $\beta$ is relatively small; the main source of the error is
the uncertainty in determining the positions of maxima and minima in the
data shown in the inset of Fig. 3(a), and the error of $\pm$0.1 in
$\beta$ is a conservative estimate. The obtained $\beta$ of 0.4 $\pm$
0.1 is reasonably close to the value $\beta$ = 0.5 expected for massless
Dirac Fermions, which points to the topological SS origin of the SdH
oscillations. Using the Onsager relation $F$ = ($\hbar$$c$/2$\pi$e)$A$,
where $A$ is the extremal FS cross-section area, we find the Fermi wave
vector $k_{\rm F}$ = 5.9 $\times$ 10$^{6}$ cm$^{-1}$, which corresponds
to the surface carrier density $n_{s}$ = 2.8 $\times$ 10$^{12}$
cm$^{-2}$ for a spin-nondegenerate surface state. It is worth noting
that if one assumes that the SdH oscillations originate from a bulk FS,
the bulk carrier density implied by $F$ is of the order of 10$^{18}$
cm$^{-3}$, which is totally inconsistent with the large $R_H$ value
observed at 1.8 K ($\sim$ --100 cm$^{3}$/C). Hence, one can conclude
that the SdH oscillations are certainly coming from the surface. From
the obtained $k_{\rm F}$ value and the Fermi velocity $v_{\rm F}$
$\simeq$ 4.6 $\times$ 10$^{5}$ m/s,\cite{RenBTS} $E_{\rm F}$ is
estimated to be $\sim$170 meV above the Dirac point, pointing to the
electron character of the surface carriers.

By fitting the temperature dependence of the oscillation amplitude with
the standard Lifshitz-Kosevich (LK) theory [inset of Fig. 3(c)], we
obtained the cyclotron mass $m_{c}$ = 0.13$m_{e}$, where $m_{e}$ is the
free electron mass. Once $m_{c}$ is known, the Dingle plot [Fig. 3(c)]
gives the Dingle temperature $T_{D}$ = 12.5 K, which corresponds to the
surface quantum mobility $\mu_s^{\rm SdH}$ of $\sim$1300 cm$^{2}$/Vs.
This $\mu_s^{\rm SdH}$ value is roughly twice as large as that reported
for thick BTS$_{1.05}$ samples.\cite{RenBTS} Surprisingly, according to
these results, the ratio of the estimated surface conductance to the
total conductance, given by $G_{\rm s}/{\sigma}t$, is calculated to be
about 1.8 in this sample; here, the surface sheet conductance $G_s = e
n_s \mu_s^{\rm SdH}$ $\simeq$ 5.8 $\times$ 10$^{-4}$ $\Omega^{-1}$, the
measured overall conductivity $\sigma$ was 0.64 $\Omega^{-1}$cm$^{-1}$
at 1.6 K, and $t$ = 6 $\mu$m.\cite{note} Obviously, it is unphysical to
have $G_{\rm s}/{\sigma}t >$ 1, and this result implies that the actual
surface transport is hindered by steps between terraces created upon
cleaving, giving the effective surface conductance that is smaller than
the intrinsic one given by $n_s$ and $\mu_s$. Note that a similar
problem was reported in Ref. \onlinecite{BTSberryphase}. In any case,
the SdH oscillations give direct estimate of the surface transport
parameters, which indicate that a surface-dominated transport can be
achieved in Sn-doped BTS crystals with several $\mu$m thickness. The
remaining bulk contribution is most likely the degenerate transport
through the impurity band which is located in the bulk band gap; to
further reduce the bulk contribution, one needs to achieve the Anderson
localization in this impurity-band transport.

\begin{figure}
\includegraphics*[width=8cm]{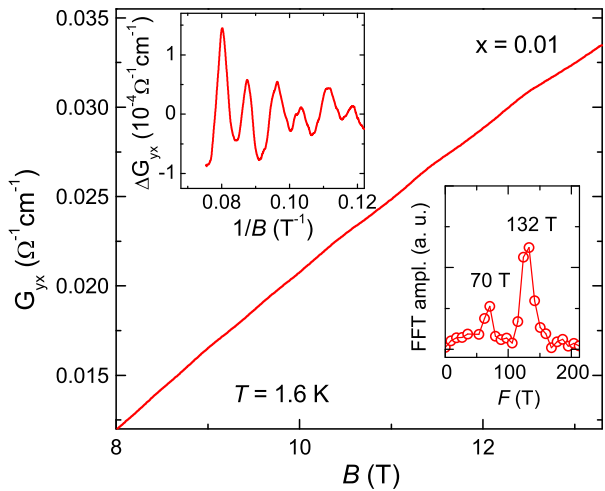}
\caption{(Color online)
$G_{yx}$ for a sample with $x$ = 0.01 at 1.6 K plotted as a function of
$B$ applied along the $C_{\rm 3}$
axis. The upper inset shows the oscillatory component
$\Delta$$G_{yx}$ plotted as a function of $1/B$. The lower inset
shows the Four transform of $\Delta G_{yx}(B^{-1})$, revealing two prominent
frequencies.
}
\label{fig4}
\end{figure}

Apart from this simplest case, the patterns of the observed SdH
oscillations were complicated. One of such examples, which was observed
in a sample with $x$ = 0.01, is shown in Fig. 4. As can be seen in the
lower inset, the Fourier transform of $\Delta$$G_{yx}(B^{-1})$ reveals
two well-resolved peaks at $F_{1}$ = 70 T and $F_{2}$ = 132 T,
reflecting the beating visible in the oscillation data (upper inset).
The $k_{\rm F}$ values calculated from the Onsager relation are 4.6
$\times$ 10$^{6}$ and 6.3 $\times$ 10$^{6}$ cm$^{-1}$ for $F_{1}$ and
$F_{2}$, respectively. These correspond to $n_{s}$ of 1.7 $\times$
10$^{12}$ and 3.2 $\times$ 10$^{12}$ cm$^{-2}$, respectively.
Unfortunately, as was pointed out in Ref. \onlinecite{RenBi2Se3}, the
multi-component nature of the oscillations prevents us from reliably
extracting the cyclotron mass or the Dingle temperature for each
component. Also, the Landau-level fan diagram is not very reliable for
extracting $\beta$ in the multi-component case.

\section{Discussions}

From the above results, it is clear that Sn acts as an acceptor in BTS.
Ideally, one would expect that the total number of holes introduced by
Sn doping, $\Delta$$n_{h}$, increases linearly with increasing $x$.
However, as can be seen in Fig. 2(d), $\Delta$$n_{h}$ becomes nearly
constant for $x$ $\geq$ 0.004, which deviates significantly from the
ideal situation. Therefore, there must be some additional effects that
lead to deactivation of Sn dopants. In this respect, a similar problem
was noted in previous studies of Sn doping in Bi$_{2}$Te$_{3}$,
\cite{HeavilySnBTS, SnBTS-Kulbachiskii, SnBTS-resonant} and two
possibilities have been proposed to explain the apparent discrepancy.

One possibility considered for Bi$_{2}$Te$_{3}$ is that a part of the Sn
atoms are built into the lattice in such a way that a seven-layer
lamellar structure, $i.e.$ Te-Bi-Te-Sn-Te-Bi-Te, is
formed.\cite{HeavilySnBTS} If a similar structure exists in Sn-doped BTS
as well, it would be Se-Bi-Te-Sn-Te-Bi-Se. It is important to note that,
in contrast to those occupying the Bi site, the Sn atoms in this
structure do not bring any charge to the lattice, and thus the
discrepancy is reconciled. However, no additional diffraction peaks
except for those corresponding to the chalcogen-ordered tetradymite
structure were observed in the XRD data even for the sample with the
highest Sn-doping concentration (data not shown), suggesting that the
seven-layer lamellae, if exist, are randomly distributed in the lattice.

Another, more plausible possibility is that, at high Sn-doping
concentration, the wavefunctions of the Sn acceptors overlap
significantly, leading to the formation of an impurity band
(IB).\cite{SnBTS-Kulbachiskii,SnBTS-resonant} Within this picture,
increasing Sn content results in an increase in the density of states
(DOS) of the IB instead of hole doping. In Sn-doped Bi$_{2}$Te$_{3}$,
this Sn-related IB was shown to be located at 15 meV below the top of
the upper valence band.\cite{SnBTS-Kulbachiskii} However, in the case of
Sn-doped BTS, this Sn-related IB is most likely located within the bulk
band gap, because the activation energy was found to be much larger in
Sn-doped BTS than in BTS$_{1.05}$, which naturally points to the
appearance of a new IB to pin the chemical potential in Sn-doped BTS.
Note that, while there may also be IBs due to Se vacancies and Bi/Te
antisite defects in Sn-doped BTS, it is most likely that $E_{\rm F}$ is
pinned to the Sn-related IB due to its large DOS. The fact that the
activation energy of $\sim$120 meV is essentially unchanged for a range
of Sn concentrations ($x$ = 0.004--0.02) is also consistent with this
picture. Further studies are called for to clarify the details of the
IBs in Sn-doped BTS.

\section{Conclusion}

We performed a systematic study of the transport
properties of a series of Bi$_{2-x}$Sn$_{x}$Te$_{2}$Se single crystals
with 0 $\leq$ $x$ $\leq$ 0.02. It is found that Sn behaves as an
acceptor, which enables us to tune the Fermi level that is located in
the conduction band in the undoped stoichiometric compound. For $\geq$
0.004, $E_{\rm F}$ is successfully tuned into the bulk band gap, and the
resistivity becomes as large as several $\Omega$cm at low temperatures.
The transport properties at high temperatures show a thermally
activated behavior with a large activation gap, which is probably
related to the formation of a Sn-related impurity band. The analysis of
the SdH oscillations observed in a 6-$\mu$m thick sample indicates
that a surface-dominated transport can be achieved in
Sn-doped BTS single crystals with several $\mu$m thickness.
This, along with the large activation gap,
makes the Sn-doped BTS system well suited for future applications of
topological insulators.

\begin{acknowledgments}
This work was supported by JSPS (NEXT Program), MEXT (Innovative Area
``Topological Quantum Phenomena" KAKENHI 22103004), and AFOSR (AOARD
104103 and 124038).
\end{acknowledgments}


\begin{thebibliography}{99}
\bibitem{footnote}
In Bi$_{1.5}$Sb$_{0.5}$Te$_{1.7}$Se$_{1.3}$ samples with the thickness
of several $\mu$m, we observed \cite{BSTStaskin} that the
low-temperature $R_{\rm H}$ undergoes a sign change with time and that
the SdH frequency accordingly shows a drastic change. However, for the
present sample, only a slight time evolution in $R_{\rm H}$ and the SdH
frequency was observed.
\bibitem{note}
It should be noted that in the present experiment, the bottom surface of
the sample was glued to a Kapton tape, so the top and bottom surfaces
were in different environments. This means that the surface chemical
potentials (and hence the surface carrier densities) for the top and
bottom surfaces were most likely different; nevertheless, only a single
frequency was observed in the SdH oscillations, suggesting that only one
surface was responsible for the oscillations. We therefore considered
only one surface in this calculation of $G_s$.
\end{thebibliography}
\end{document}